\documentclass[aps,prl,twocolumn,10 pt,groupaddress,floats,showpacs]{revtex4}
\usepackage{graphicx}
\begin{document}
\title{Electrodynamics of the vortex lattice in untwinned YBaCuO by complex impedance measurements}
\author{A. Pautrat, A. Daignere, C. Goupil, Ch. Simon}
\address{CRISMAT/ISMRA, UMR 6508 associée au CNRS, Bd Maréchal Juin, 14050 Caen c$\acute{e}$dex, France.}
\author{B.Andrzejewski}
\address{Institute of molecular physics, Smoluchowskiego 17, Poznan, Poland.}
\author{A.I. Rykov, S. Tajima}
\address{Superconductivity Research Laboratory, ISTEC, Tokyo 135, Japan.}
\date{\today}

\begin{abstract}

We report complex impedance measurements in an untwinned YBaCuO crystal. Our broad
frequency range covers both the quasi static response and the resistive response of the
vortex lattice. It allow us to characterize the irreversibility line without the need of
any frequency dependent pinning parameters. We confirm the validity of the two modes
model of vortex dynamic, and extract both the surface critical current and the flux flow
resistivity around the first order transition $T_{m}$. This latter is identified by the
abrupt loss of pinning and by an unexpected step of $\rho_{ff}(T)$ at $T_{m}$.
\end{abstract}

\newpage
\maketitle
 The dynamic response of a vortex lattice is mainly determined by the critical current $I_{c}$ and by the flux flow resistivity $\rho_{ff}$.
 The classical method used to measure these values is to pass a transport current in
the superconducting sample and to measure the voltage $V$ generated by the vortex lattice
flow. A critical current can be extracted by extrapolating the linear part of this $V(I)$
curve down to $V=0$, and the flux flow resistance is given by the slope of this linear
part. This simple and powerful method is unfortunately not applicable for high $T_{c}$
materials, due to the high critical current in the pinned vortex state, and to the
corresponding heating due to Joule effect in the contact leads. Another way to proceed is
to measure the high frequency skin depth of the complex penetration \cite{nonoupt3}. This
response has been first observed by Gittleman and Rosenblum, who noticed that the linear
high frequency response of a pinned vortex lattice mimics the ideal response of a free
vortex lattice (the viscous force becomes greater than the "pinning force" at high
frequencies) \cite{Gittleman}. On the contrary, the low frequency regime is linked to the
critical current via the quasi static regime (the "Campbell" regime) where the
penetration $\lambda_{ac}$ is purely real (inductive response) as in the Meissner state
\cite{campbell}. Between these two regimes, the frequency spectrum around the depinning
frequency $\Omega_{p}$ is the quantity of interest for discriminating between different
types of pinning state (bulk or surface)\cite{nono}. Therefore, the frequency
 response of the vortex lattice allows investigation of the inductive and the
resistive regimes. It is a powerful method to characterize both pinned and depinned
vortex states. A small ac field is used as a probe, and one detects the vortex response
in the form of a susceptibility, surface impedance or resistivity. $HT_{c}$ materials
have been extensively studied by these different ac techniques, but it is quite difficult
to draw an unique picture. Ac susceptibility focuses mainly on the so called loss peak
($\chi_{"}$ peak) and on its frequency dependence. After strong controversy about its
significance (onset of superconductivity \cite{worthington}, melting transition
\cite{gammel}, depinning transition \cite{vanderberg}...), other interpretations deal
with the finite size effect of a resistivity driven by thermal depinning \cite{steel}
\cite{supple}. On the other hand, strong contradictions exist between the resistivity
values and the expected geometrical frequencies \cite{steel}, which raise questions about
this theoretical treatment of vortex dynamics.

Among other cuprates, the case of untwinned YBaCuO is of particular interest. It is now
obtained with a large enough size to avoid spurious size effects which may obscure
experimental signatures contained in the full spectrum of the depinning transition.
Moreover, thermodynamical evidence of a first order transition separating a pinned vortex
state and a depinned vortex state have appeared recently \cite{marcenat}. This transition
is usually interpreted as the melting of a Bragg-Glass phase into a liquid phase without
pinning \cite{thierry}. By analogy with real crystals, thermally induced displacements of
bulk pinned vortices added to a Lindemann criterion are the key elements of this
transition. Nevertheless, the properties of these two phases are not so well known. In
particular, it is usually assumed that vortex lattice pinning and dynamics are quite
different in high $T_{c}$ materials compared to what is currently observed in low
$T_{c}'s$. To evidence peculiarities of vortex dynamics in $HT_{c}$ materials, one has to
perform experiments, the analysis of which successfully applies in conventional
superconductors. As an example, high temperatures should lead to strong thermally
activated behaviour, which is supposed to be reflected in pinning properties, leading to
dominant thermally driven depinning. In such a case, the ac response should be quite
different from what is observed in the samples where thermal activation was shown to be
negligible for vortex depinning \cite{campbellevetts}.

In this paper, we present a study of vortex lattice depinning in an untwinned YBaCuO
crystal by mean of complex surface impedance measurements. Due to the important role of
skin effects, we express this impedance in the form of a complex penetration depth.
 If the measurements of these complex penetration depths
versus temperature looks quite complicated to interpret, due to frequency dependent
features, we show that a study of the depinning spectrum leads to a simpler picture of a
surface pinned vortex lattice with a bulk free flow resistivity, as it is observed in
several conventional low $T_{c}$ superconductors \cite{nono}. Nevertheless, the data
exhibit two important differences. We observe the well known disappearance of the
critical current at $T_{m}<T_{B_{c2}}$, but this disappearance is simultaneous with a
less expected step in the temperature dependance of $\rho_{ff}$.

\begin{figure}[h]
\begin{center}
\vspace{0.5cm} \hspace{-0.3cm} \resizebox{!}{0.2\textwidth}{\includegraphics{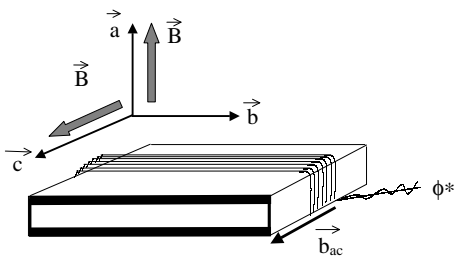}}
\caption{Geometry of the experiment. The quantity of interest is the measured flux
$\phi_{ac}=\int b_{ac} dS\approx2\lambda_{ac}\ell_{b}b_{o}$ in a $1D$ approximation.}
\end{center}
\end{figure}

 The sample is an untwinned crystal of $YBa_{2}Cu_{3}O_{7-\delta}$ ($(\ell_{a}=
L= 600, \ell_{b}= 3000,\ell_{c}=1000)\mu$m$^{3}$), the preparation of which was detailed
in ref \cite{rykov}. The post annealing procedure and the high $T_{c}$ of $93.5 K$
corresponds to the optimally doped state with $\delta=0.07$ in the Lindemer scale. Normal
state resistivitity $\rho_{b}$ has been measured using standard four probe techniques and
gives a value of about 40  $\mu\Omega.cm$ at $100 K$.  Both this low value and the high
$T_{c}$ attest to the good quality of the sample, as do the observed magnetization step
at the first order transition \cite{rykov2}. The main part of the experimental set up
consists of a waveform generator (DS345) and two lock-in amplifiers (SR850 and SR844), so
as to cover a frequency range of about $30 Hz- 30 MHz$. The ac response is the flux taken
by a small pick-up coil, directly glued to the sample, which reposes, itself, in the
excitation coil. We have carefully checked that the applied alternative field
$b_{ac}=b_{o}exp(-i\Omega t)$ has a low enough magnitude $(\approx 1 \mu T)$ to stay in
the linear regime. The complex penetration depth is then given by $\lambda _{ac}=\phi
_{ac}/2 \ell_{b} b_{o} =\lambda ^{\prime }+i\lambda "$ (Fig. 1). The whole set up drives
to an experimental resolution of few microns, and hence the London penetration can be
easily neglected. Therefore, the calibration of the phase and of the zero of penetration
have been done using the Meissner state as a reference ($\lambda^{'}=\lambda^{"}=0$). For
the highest frequency points, we have renormalized the signal using a small reference
coil near the sample (when high frequencies and circuitry began to cause phase shifts).
The complete penetration has been measured in the normal state at 100 K and at low
frequency and the resistivity value has been confirmed by a skin effect fit which gave a
value of $\rho_{b}\simeq 40.5 \mu\Omega.cm$. As the theoretical models of interest are
one dimensional in the simplest case, one has to choose a geometry which is as close as
possible to a 1D penetration (Fig. 1). $b_{ac}$ is applied along the c-axis and the flux
is measured through the ($\overrightarrow{a},\overrightarrow{b}$) surface. Most of the
data were taken in the geometry $\overrightarrow{B}//\overrightarrow{a}$. In this case,
the currents are mainly confined in the ($\overrightarrow{b},\overrightarrow{c}$)
surfaces. This is specially true at high frequencies which restrict the penetration of
the resistive wave in a thin layer. Vortices are shaken on the
($\overrightarrow{b},\overrightarrow{c}$) surface and the wave penetrates along the
$\overrightarrow{a}$ direction. In the mixed state, there is also an anisotropy induced
by the Josephson relation
($\overrightarrow{E}=-\overrightarrow{V_{L}}\wedge\overrightarrow{\omega}$ so the
electric field // $\overrightarrow{j_{b}}$ is perpendicular to the vortex field
$\overrightarrow{\omega}$). The small part of the current along the $\overrightarrow{a}$
direction (along the vortices) can thus be neglected. To confirm the main conclusion of
this study, a few measurements have been taken with
$\overrightarrow{B}//\overrightarrow{c}$. Even if this geometry does not allow as many
quantitative results, because of the problems of the closing currents along the
$\overrightarrow{a}$ direction which are now perpendicular to vortices, the results were
typically the same. The data presented here were taken at a magnetic field of 6 T.

\begin{figure}[h]
\begin{center}
\vspace{0.5cm} \hspace{-0.3cm} \resizebox{!}{0.4\textwidth}{\includegraphics{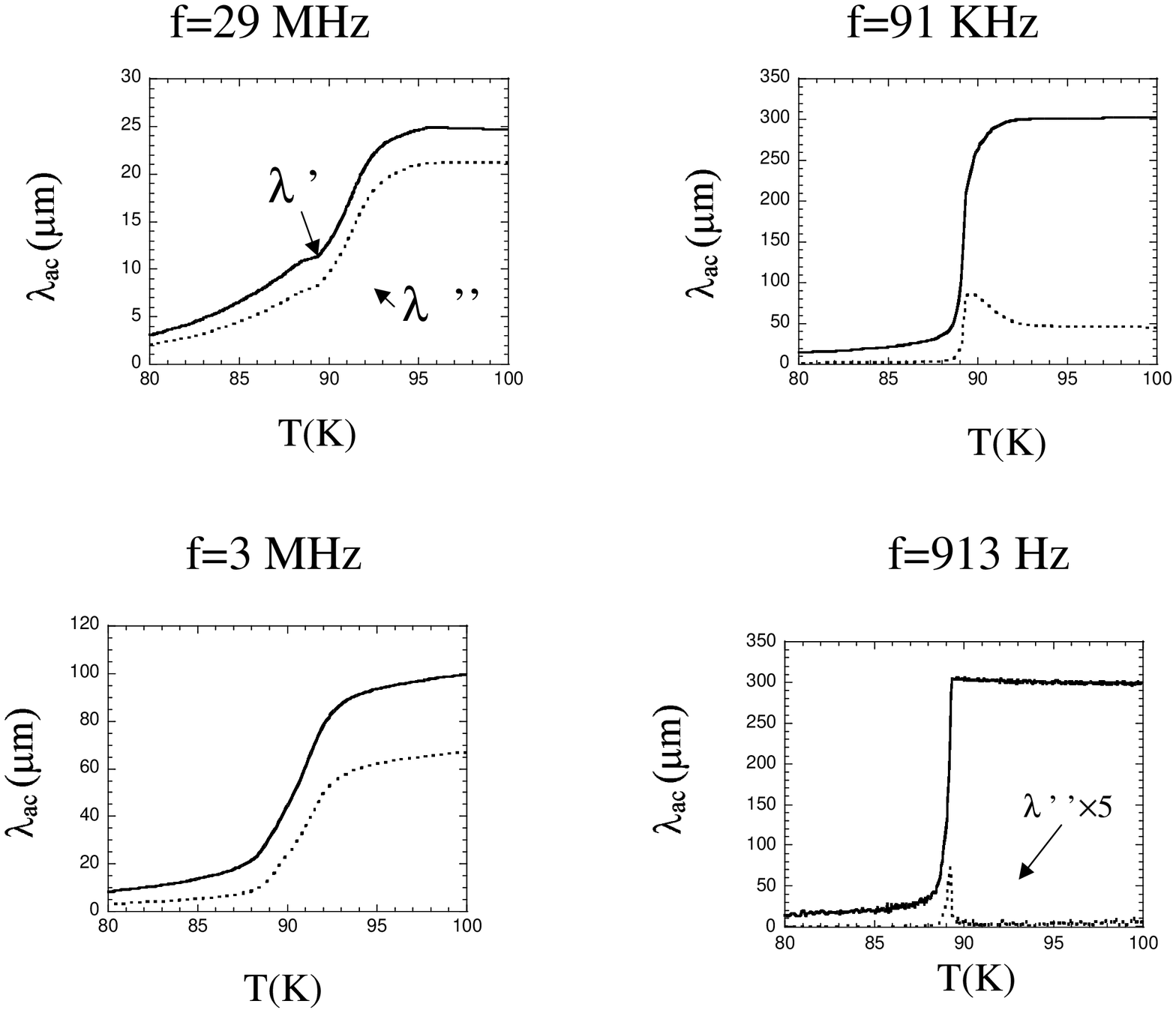}}
\caption{$\lambda_{ac}$ as function of the temperature for a fixed field $(B//a)=6T$
($\lambda^{'}$ in full lines, $\lambda^{''}$ in dotted lines). Note the change of scale
of penetration length for the different frequencies. One can evidence both the inductive
transition (low frequency, arctan$\frac{\lambda^{''}}{\lambda^{'}}\approx0 deg$ in the
pinned state, full penetration in the depinned state) which probes the pinning response,
and the quasi-resistive response (high frequency,
arctan$\frac{\lambda^{''}}{\lambda^{'}}\approx\pi/4$).}
\end{center}
\end{figure}

Before discussing the experimental results, we have to recall the main features of the
linear response of a vortex lattice. At high frequencies $\Omega \gg \Omega_{p}$, all the
models of ac response predict the response of a resistive medium:
$\lambda_{ac}=\lambda_{bulk}=\lambda_{ff}=\frac{(1+i)\delta_{ff}}{2}$, where
$\delta_{ff}=\sqrt{\frac{2 \rho_{ff}}{\mu_{o}\Omega}}$ is the usual flux-flow skin depth.
A review of the ac response of bulk pinned vortices can be found in the ref
\cite{vanderbeek}. The main idea is that the bulk pinning response is governed by a
modified skin depth equation with one mode. In the simplest case, this leads to
$K_{bulk}=(\lambda_{c}+\frac{\delta_{ff}}{2i})^{-1}$ \cite{campbell}, ($\lambda
_c=\sqrt{\frac{B^{2}}{\mu_{o}}\alpha_{Lo}}$ is the campbell length and $\alpha _{Lo}$ the
Labusch parameter ). The thermal activation is taken into account when rewriting
$\alpha_{Lo}$ as $\alpha_{L}=\alpha_{Lo}\frac{i\Omega}{1+i\Omega\tau}$ where
$\tau=\tau_{o}exp(\frac{U}{kT})$ is the creep relaxation time \cite{vanderbeek}
\cite{coffey} \cite{brandt}. It is formally equivalent to introducing a low frequency
skin effect governed by an activated resistivity $\rho_{ff}exp-(\frac{U}{kT})$. Bulk
pinning models with thermal activation predicts so two depinning frequency,
$\Omega_{p}=\frac{\rho_{ff}}{\mu_{o}\lambda_{c}^{2}}$ and
$\widetilde{\Omega_{p}}=\Omega_{p} exp -(\frac{U}{kT})$. The low frequency response
$(\Omega\rightarrow0)$ must follow the law $\lambda_{ac}\propto\Omega^{-\frac{1}{2}}$
(resistive response). The ac frequency spectrum should allow the extraction of the
Boltzman like factor $exp -(\frac{U}{kT})$. We emphasize that this kind of resistive
response is at odds with the response in a Campbell-like regime
$\lambda_{ac}\simeq\lambda^{'}\simeq cte$ (inductive response) and is easily identifiable
if one measures the phase of $\lambda_{ac}$.

It was also shown, both theoretically \cite{Sonin}\cite{placais} and experimentally
\cite{nono}, that, taking into account vortex elasticity and appropriate boundary
conditions for vortex lines, a non dissipative penetration mode
$k_{surf}=\frac{i}{\lambda_{surf}}$ adds in the dispersion equation. This mode has a
short spatial scale but, as shown in reference \cite{MS}, allows for strong curvature of
vortex lines. This mode is of particular importance for treating the case of surface
pinning. Its weight is enhanced by the surface roughness present in any real sample and
allows for the flow of a large non dissipative current (the critical current). Taking
into account the finite size of the sample, the complex penetration depth was shown to
take the form:

\begin{equation}\label{surfsize}
\frac{1}{\lambda_{\rm ac}}=\frac{1}{L_{\rm S}}+ \frac{1-\mathrm{i}}{\delta_{\rm
ff}}\mathrm{cotanh} \frac{(1-\mathrm{i})L}{2\delta_{\rm ff}}  \qquad.
\end{equation}

 where $\delta_{ff}$
is the only frequency-dependent parameter.
$L_{s}\approx\lambda^{'}(0)\approx\frac{a_{o}\omega}{\mu_{o}i_{c}}$ is related to the
superficial critical current $i_{c}$ $(A/m)$ and $L$ is the thickness of the sample
(L=$\ell_{a}$ in our geometry). Note that $L_{s}$, as $i_{c}$, is independent of the
frequency (as long as $\Omega \ll \Omega_{gap}$). The bulk pinning contribution can be
simply introduced following the same approach as the Campbell one. This results in a
strong narrowing of the depinning spectrum, easily observable by experiment \cite{nono}.
Since the spectra measured in the present work always follow equation (1) with an
accuracy within the noise of the data. We reach the conclusion that
$\lambda_{c}^{-1}\approx0$, i.e. bulk pinning is negligible compared to surface pinning
in this sample.

\begin{figure}[h]
\begin{center}
\vspace{2.5cm} \hspace{1.5cm} \resizebox{!}{0.2\textwidth}{\includegraphics{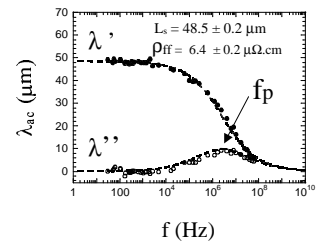}}
\caption{Depinning spectrum of the pinned vortex lattice ($T=88.6 K < T_{m}$). The dashed
line is a fit using equation(1) (pure surface pinning with $L_s$=48.5 $\mu$m and
$\rho_{ff}$=6.4 $\mu\Omega$.cm ). Other possible bulk pinning contribution is found
negligible.}
\end{center}
\end{figure}

 Some typical penetration depths $\lambda_{ac}(T)$, measured for different frequencies and
at a fixed field B=6T, are shown in figure 2. We note the maximum of the out of phase
component $\lambda^{"}$ and the saturation of $\lambda^{'}$ for frequency dependent
temperature values. This can be associated with a thermal depinning of the vortex
lattice. Anyway, because of the possible mixing of many effects, both intrinsic (pinning,
flow, transition of the vortex lattice...) and extrinsic (finite size effects) and of the
bad known temperature variation of thermodynamic parameters of the vortex lattice, a
precise study of the ac dynamical response needs a study of the frequency dependence.
Such a spectrum taken in the "solid" phase $(I_{c}\neq0)$ is presented in the figure 3.
It reveals a two mode spectrum, with a pure surface pinning and free vortex flow in the
bulk, as previously observed in conventional superconductors \cite{nono}, and in a
slightly overdoped YBaCuO \cite{nous}. Other spectra taken at lower temperatures confirm
this result (Fig.4).

\begin{figure}[h]
\begin{center}
\vspace{0.5cm} \hspace{-0.3cm} \resizebox{!}{0.2\textwidth}{\includegraphics{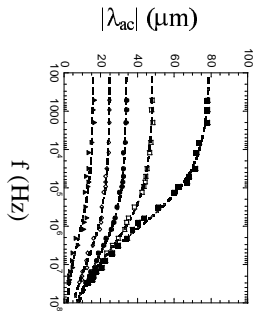}}
\caption{Frequency spectra of the vortex lattice in the "solid" state($I_{c}\neq0$, T=
78.4, 85.7, 87.8, 88.5, 88.8 K). The low frequency penetration increases (and the surface
critical current decreases) as the temperature increases. Note the broad depinning
spectrum due to the surface mode compared to the one mode skin effect spectrum of fig 6.
The dashed line is a fit with equation (1) with the following parameters: $L_s$=15.8,
24.7, 34.1, 48.5, 79.6 $\mu$m and $\rho_{ff}$=0.3, 3.1, 6.2, 6.5, 6.4 $\mu\Omega$.cm. }
\end{center}
\end{figure}

\begin{figure}[h]
\begin{center}
\vspace{4.5cm} \hspace{-0.3cm} \resizebox{!}{0.2\textwidth}{\includegraphics{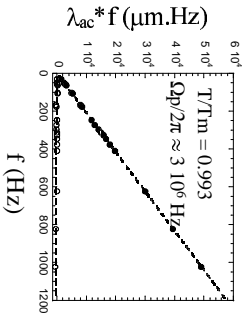}}
\caption{The low frequency response of the vortex lattice $\lambda_{ac} \times$ $f$ as
function of the frequency $f$, so as to emphasize the low frequency linearity which
implies $\lambda^{'}\simeq 48.5 \mu$ m $< \ell_{a}/2$.}
\end{center}
\end{figure}

Comparing with previous measurements \cite{nous}, we have extended the frequency range so
as to investigate the pure loss free response and the pure resistive response in the
overdamped regime. In particular, the low frequency behaviour has been measured near
$T_{m}$ ($T/T_{m}\simeq0.993$), in order to increase the probability of thermally
activated vortex jumps. As evidenced in figure 5, $\lambda'$ is nevertheless constant,
within noise due to the smallest of low frequency signal, and there is no loss
($\lambda^{''}=0$) down to a frequency of about $10^{-5}\Omega_{p}$. In others words, the
critical current does not depend on the frequency even very close to $T_{m}$, and creep
effects are thus negligible up to our lowest measurable frequency (few tens Hz) even
close to $T_m$.

\begin{figure}[h]
\begin{center}
\vspace{0.5cm} \hspace{-0.3cm} \resizebox{!}{0.2\textwidth}{\includegraphics{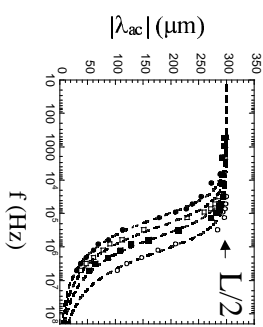}}
\caption{Frequency spectra of the vortex lattice in the "liquid" state($I_{c}=0$, T=
89.3, 89.9, 90.9, 91.9 K). The low frequency depth is a constant and given by the half
thickness of the sample. The shift of the skin effect frequency is due to the growing
flux flow resistivity. The dashed line is a fit with equation(1) with the following
parameters: $L_{s}^{-1}\propto i_{c}=0$ and $\rho_{ff}$=6.9, 9.9, 14.3, 29.5
$\mu\Omega$.cm. }
\end{center}
\end{figure}

 In the "liquid" phase $(T>T_{m}, I_{c}=0)$,
one measures the spectrum of a superconductor free from defects, as expected from the
disappearance of any vortex pinning (fig 6). This is a classical electromagnetic skin
effect driven by the flow resistivity $\rho_{f}$, growing up to $T_{BC2}$. As there is no
more critical current, the surface mode is turned off ($L_{s}^{-1}=0$ in the equation
(1)) and only the bulk resistive mode is active. The low frequency penetration is that of
a  perfectly transparent slab: $2L=\lambda _{ac}$, which allows identification of the
irreversibility temperature without ambiguity. Let us emphasize that $2L=\lambda
_{ac}(\frac{\Omega}{2\pi} \rightarrow 0)$ only for $T \geq T_{m}=89.2 K$, and that
$T_{m}$ is not frequency dependent. As the critical current and the irreversibility line
do not depend on the frequency, the pinning parameters are not dispersing in the
frequency range we have investigated. We conclude that we do not observe any role of
thermal assisted depinning of vortices.

 In figure 7, we present the two parameters extracted from the low frequency and the high frequency curves, the
critical current and the flux flow resistivity. In the surface pinning model, $i_{c}$ is
simply link with the main reversible magnetization $M=-\int_{vol}\varepsilon dV$, as
diamagnetic currents and surface non dissipative currents are the same kind of
equilibrium currents \cite{MS}. The only adjustable parameter is the phenomenological
angle $\theta_{cr}\simeq\frac{i_{c}}{\varepsilon}$, which takes into account both the
surface roughness and the vortex lattice elasticity. We have measured the main reversible
magnetization of the sample by means of a squid magnetometer. Comparing with $i_{c}$
($T<T_{m}$), this leads to a reasonable and quite standard value of $\theta_{cr}\approx5
deg$ (corresponding to a standard surface roughness) which goes to zero at $T= T_{m}$.

\begin{figure}[h]
\begin{center}
\vspace{0.5cm} \hspace{-0.3cm} \resizebox{!}{0.2\textwidth}{\includegraphics{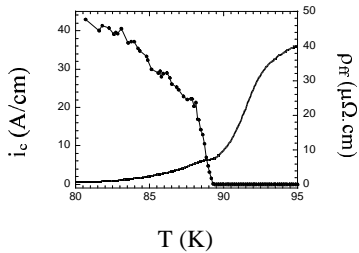}}
\caption{The surface critical current and the flux flow resistivity around the first
order transition.}
\end{center}
\end{figure}

We will now focus on the cross over between the pinned and depinned state. As now well
established, some very clean YBaCuO samples present a thermodynamic first order
transition \cite{marcenat}, at a field (temperature) where the critical current
disappears. Such a feature is usually interpreted as the melting of a vortex solid into a
liquid state. The vortex solid is Abrikosov-like, and more recent theories predict a
quasi long range ordered lattice in the case of low enough pinning (the Bragg Glass)
\cite{thierry}. The properties of the "liquid" state are badly known. As structural
studies of the high temperature state of the VL are not possible due to the lack of
resolution so close to $T_c$, most of the experimental answers are indirect. One can use
transport measurements in order to test a pertinent property. Recent complex resistivity
measurements in YBaCuO have interpreted the inductive to resistive transition as a
collapse of the vortex shear constant $C_{66}$ \cite{mapple}, as claimed if the first
order transition is a genuine melting. We have shown here and discussed in
\cite{comment}, that this transition is simply described by the disappearance of the
pinning strength, as it appends at $B_{C2 }$ in low Tc materials. However, another
feature appears in our data. The high frequency $\lambda_{ac}(T)$ curves present a step
at the temperature of the first order transition. This is observed in the two directions
(B//a-axis and B//c-axis). The remaining question is what could cause this feature. In
principle, when measuring at high enough frequencies, the vortex lattice behaves as if it
was ideal leading to $\lambda_{ac}\simeq\lambda_{ff}=\frac{(1+i)\delta_{ff}}{2}$. In
fact, one has to correct this result by the diamagnetism of the mixed state
$\mu=\mu_{o}\frac{B}{B+\mu_{o}\varepsilon}$,i.e
$\lambda_{ac}\simeq\frac{\mu}{\mu_{o}}\lambda_{f}$ and
$\delta_{ff}=\sqrt{\frac{2\rho_{ff}}{\mu\Omega}}$ \cite{placais}. As the first order
transition suddenly modifies this diamagnetism by $\Delta\varepsilon$, it could lead to a
concomitant effect on $\lambda_{ac}$. Nevertheless, as
$B\gg\mu_{o}\Delta\varepsilon(\lesssim10^{-4} T)$, this effect is far too small to affect
our measurement. Moreover, at the highest frequencies used, $\delta_{ff}\lesssim 50
\mu$m$<L/10$, so the thick limit is well justified and spurious size effects are
neglected. We have also checked that the values extracted both from a fit of the complete
spectra or at a constant frequency f $\gg$ $f_p$ are equivalent. Our conclusion is that
the step of $\lambda_{ac}$ reflects the one of $\rho_{ff}$. We emphasize that in
conventional superconductors, there is no link between the flux flow resistivity and the
critical current \cite{Rose}, except at a critical field as $B_{C2}$ ($I_{c}\approx0$,
$\rho_{ff}\approx\rho_{n}$). Such independence between the pinning characteristics and
the bulk flow resistivity of a superconductor in the mixed state was evidenced in ref
\cite{rosenblum}. As an example, the flux flow resistivity, measured in dc or at high
frequencies, was shown to be the same and clearly insensitive to the critical current
peak effect in conventional superconductors \cite{kim}. On the contrary, we see here that
the first order transition affects both the critical current and the flux flow
resistivity. As this latter is linked to the nature of the order parameter and to the
quasiparticle excitations in the vortex state, it is tempting to conclude that the first
order transition affects the electronic structure inside and (or) around the vortex core,
or in contrast that a transition in this electronic structure drives the first order
transition. The difficulty lies in understanding the key role of the different
peculiarities of the vortex core in YBaCuO (short coherence length, d-wave symetry with
nodes, strong antiferromagnetic fluctuations even for optimally doped samples
\cite{veran}) compared to the simplest case of a conventional dirty s-wave
superconductor. A question of the same type has been previously addressed by D'anna et al
\cite{Danna}, who observed a strong decrease of the Hall resistivity at a temperature
just below the one of the first order transition. We note that a small distortion of the
vortex lattice in a d-wave superconductor should lead to an important thermodynamic
contribution ($\Delta$S$\approx$k$_{B}$) via the fermionic entropy \cite{Volovik}, and
that such kind of distortion does not necessary correspond to a melting.

\begin{figure}[h]
\begin{center} \vspace{2.5cm} \hspace{2.5cm}
\resizebox{!}{0.2\textwidth}{\includegraphics{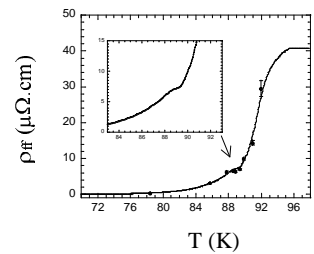}} \caption{flux flow resistivity
extracted from a fit of the whole depinning spectrum (point) compared to the values
measured at a constant frequency ($f=29 MHz$) assuming
$\lambda_{ac}=\lambda_{ff}=\frac{(1+i)\delta_{ff}}{2}$ (solid line).}
\end{center}
\end{figure}

In conclusion, complex impedance measurements have been performed in a clean YBaCuO
crystal, with a particular attention to the vortex depinning transition. Two mode
electrodynamic analysis of the data allow us to conclude that the vortex pinning is due
to the surface roughness of the sample, and that the depinning (and the irreversibility
line) is not driven by thermal activation. The flux flow resistivity exhibits a step at
the first order transition, that remains to be interpreted, in the framework of a melting
or using alternative descriptions.

This work was partially supported by NEDO, Japan for the R and D of Industrial Science
and Technology Frontier Program.

AP, CG and CS are indebted to Bernard Pla\c{c}ais et al (ENS, PARIS) whose work is at the
origin of this type of experiment.


\begin{thebibliography}{99}
\bibitem{nonoupt3} N. L$\ddot{u}$tke-Entrup, R. Blaauwgeers, B. Pla\c{c}ais, A. Huxley, S. Kambe, M. Krusius, P. Mathieu, and Y.
Simon, Phys Rev B 64, 020510 (2001).
\bibitem{Gittleman} J.I Gittleman and B. Rosenblum, Phys. Rev. Lett. 16, 734 (1966).
\bibitem{campbell} A.M. Campbell, J. Phys. C {\bf 2}, 1492 (1969).
A.M. Campbell, in Magnetic Susceptibility of Superconductors and Other Spin Systems,
edited by R.A Hein (Plenum, New York, 1992), p 129.
\bibitem{nono} N. L$\ddot{u}$tke-Entrup, B. Pla\c{c}ais, P. Mathieu, and Y.Simon, Phys Rev Lett 79, 2538 (1997); \emph{ibid}, Physica B 255, 75 (1998).
\bibitem{worthington} T.K. Worthington, W.J. Gallagher, D.L. Kaiser, F.H. Holzberg, and
T.R. Dinger, Physica C153-155, 32 (1998).
\bibitem{gammel} P.L. Gammel, L.F. Schneemeyer, J.V. Wasczak, and D.J. Bishop, Phys Rev
Lett 61, 1666 (1988).
\bibitem{vanderberg} J. Van der Berg, C.J. Van der beek, P.H. Kes, J.A. Mydosh, M.J.V.
Menken, and A.A. Menovski, Supercond Sci Technol 1, 249 (1999).
\bibitem{steel} D.G. Steel and J.M. Graybeal, Phys Rev B 45, 12643 (1992).
\bibitem{supple} F. Supple, A.M. Campbell, and J.R. Cooper, Physica C 242, 233 (1995).
\bibitem{marcenat} F. Bouquet, C. Marcenat, R. Calemczuc, A. Erb, A. Junod, M. Roulin, U.
Welp, W.K. Kwok, G.W. Crabtree, N.E. Phillips, R.A. Fisher, A. Schilling, in Physics and
Material science of vortex states, edited by R. Kossowsky (NATO science series, 1998),
p743. F. Bouquet, Phd Université de Grenoble (1998).
\bibitem{thierry} T. Giamarchi and S. Bhattacharya, "vortex phases", Lecture notes of the 2001 Cargese school on "Trends in high magnetic field science", to be published by
Springer.
\bibitem{campbellevetts} A.M. Campbell and J.E. Evetts in "critical currents in superconductors" (Taylor and Francis, London) 21, 302 (1972).
\bibitem{rykov} A.I. Rykov, S. Tajima, in "Advances in Superconductivity VIII"  (H. Hayakava et Y. Enomoto,
Springer-Verlag), 341 (1996).
\bibitem{rykov2} A. I. Rykov, S. Tajima, F. V. Kusmartsev, E. M. Forgan, and Ch. Simon
Phys Rev B 60, 7601 (1999).
\bibitem{vanderbeek} C. J. van der Beek, V. B. Geshkenbein, and V. M. Vinokur, Phys. Rev. B 48,
3393 (1993).
\bibitem{coffey} M.W. Coffey and J.R. Clem, Phys Rev B 45, 9872 (1992).
\bibitem{brandt} E.H. Brandt, Phys Rev Lett 67, 2219 (1991).
\bibitem{Sonin} E.B. Sonin, A.K. Tagantsev and K.B. Traito, Phys Rev B 46, 5830 (1992).
\bibitem{placais} B. Pla\c{c}ais, P. Mathieu, Y. Simon, E.B. Sonin, and K.B. Traito, Phys
Rev B 54, 13083 (1996).
\bibitem{MS} P. Mathieu and Y. Simon, Europhys Lett 5, 67 (1988).
\bibitem{nous} A. Pautrat, Ch. Goupil, C. Simon, N. L$\ddot{u}$tke-Entrup,
B. Pla\c{c}ais, P. Mathieu, Y. Simon, A.I. Rykov, S. Tajima, Phys Rev B 63, 054503 (2001).
\bibitem{mapple} Peter Matl, N. P. Ong, R. Gagnon, and L. Taillefer, Phys. Rev. B 65, 214514 (2002).
\bibitem{comment} A. Pautrat, C. Goupil, Ch. Simon, B. Pla\c{c}ais, P. Mathieu,
cond-mat/0207074 and accepted as a comment to Phys Rev B.
\bibitem{Rose} A.C. Rose-Innes and E.H. Rhoderick, in "Introduction to Superconductivity", second edition, (Pergamond Press)1978,
p 206.
\bibitem{rosenblum} B. Rosenblum and M. Cardona, Phys Rev Lett 12, 657 (1964).
\bibitem{kim} Y.B. Kim and M.J. Stephen in "Superconductivity" vol.2, edited by R.D. Parks, p 1120 (1969) and references herein.
\bibitem{veran} V. F. Mitrovi$\acute{c}$, E. E. Sigmund, H. N. Bachman, M. Eschrig, W. P. Halperin, A. P.
Reyes, P. Kuhns, and W. G. Moulton, Nature 413 , 505 (2001).
\bibitem{Danna} G. D'Anna, V. Berseth, L. Forro, A. Erb and E. Walker, Phys. Rev. Lett.
81, 2530 (1998).
\bibitem{Volovik} G. E. Volovik, JETP Lett. 65, 491 (1997).
\end{thebibliography}
\end{document}